Ground-level ozone following astrophysical ionizing radiation events: an additional biological hazard?

Running title: Ground-level ozone

Brian C. Thomas[1*] and Byron D. Goracke[1]

[1]Washburn University, Department of Physics and Astronomy, Topeka, KS;

[*]Corresponding author: 1700 SW College Ave., Topeka, KS 66604; 785-670-2144;

brian.thomas@washburn.edu

Abstract:

Astrophysical ionizing radiation events such as supernovae, gamma-ray bursts, and solar proton events have been recognized as a potential threat to life on Earth, primarily through depletion of stratospheric ozone and subsequent increase in solar UV radiation at Earth's surface and in the upper levels of the ocean. Other work has also considered the potential impact of nitric acid rainout, concluding that no significant threat is likely. Not yet studied to-date is the potential impact of ozone produced in the lower atmosphere following an ionizing radiation event. Ozone is a known irritant to organisms on land and in water and therefore may be a significant additional hazard. Using previously completed atmospheric chemistry modeling we have examined the amount of ozone produced in the lower atmosphere for the case of a gamma-ray burst and find that the values are too small to pose a significant additional threat to the biosphere.

These results may be extended to other ionizing radiation events, including supernovae and extreme solar proton events.

Keywords:

Supernovae, Extinctions, Ozone

1. Introduction

A variety of potential impacts to life on Earth from astrophysical ionizing radiation events have been recognized. These include: direct biological damage from redistributed radiation during the event (Martin et al., 2009; Peñate et al., 2010); heightened solar ultraviolet (UV) radiation at the ground following depletion of stratospheric ozone ($O_3$) for several years after the event (Reid and McAffee, 1978; Gehrels et al., 2003; Thomas et al., 2005; Thomas, 2009; Atri et al. 2014; Thomas et al. 2015); and nitric acid rain for months to years after the event (Melott et al., 2005; Thomas and Honeyman, 2008). An additional impact that has not yet been studied is the formation of ozone in the lower atmosphere, which may be harmful to life on Earth's surface.

Simulations of the atmospheric effects of a variety of events have been completed, including gamma-ray bursts (GRBs), supernovae (SNe), and solar proton events (SPEs). That work focused primarily on production of odd nitrogen oxides, denoted $NO_y$ (including, NO, $NO_2$, $NO_3$, and others, see Thomas et al. 2005 for more details) and subsequent depletion of stratospheric $O_3$ (Gehrels et al., 2003; Melott et al., 2005; Thomas et al., 2005, 2007, 2008, 2012, 2013; Ejzak et al., 2007). Estimates of biological impacts of heightened UV have been reported in Thomas et al. (2005, 2013, 2015) and Ejzak et al. (2007). Thomas and Honeyman (2008) and Neuenswander and Melott (2015) found that deposition of nitrate would not present a threat, and would instead provide some fertilizing effect for aquatic and land plants.

Extreme ionization events such as nearby supernovae and gamma-ray bursts lead to severe depletion of stratospheric ozone; however, the decrease in $O_3$ in the middle atmosphere leads to production of $O_3$ at lower altitudes, due to increased penetration of Solar UV radiation that is normally absorbed in the stratosphere, an effect known as "self healing," observed under modern day conditions and in modeling studies (Jackman and McPeters, 1985; Hood and McCormack, 1992; Daniel et al., 1999; Mills et al., 2008).

It is known that ozone has a detrimental effect on a wide range of organisms, due to its strong oxidizing ability.  In this work, we examined changes in low altitude ozone using previously completed atmospheric chemistry modeling, and then compared modeled levels of ozone to levels determined to be harmful to terrestrial and aquatic life.

We note that formation of low-altitude $O_3$ caused by atmospheric ionization can have effects on atmospheric dynamics and temperature (see, for instance, Calisto et al., 2011).  However, in this work we are interested only in the direct biological effects of enhanced ground-level $O_3$ and other effects are beyond the scope of this paper.

2. Methods

For this study we used previously completed atmospheric modeling reported in Thomas et al. (2005) and Ejzak et al. (2007). Those papers present results of modeling the effects of a gamma-ray burst with variation of many different parameters, including fluence, time of year of event and location of event in latitude. Modeling reported in that work was performed using the NASA Goddard Space Flight Center two-dimensional (latitude, altitude) time-dependent atmospheric chemistry and dynamics model (hereafter referred to as the "GSFC model"). The model contains 65 chemical species, and computes atmospheric constituents in a largely empirical background of solar radiation and galactic cosmic ray variations, with photodissociation included. The model includes heterogeneous processes (e.g. on polar stratospheric clouds), which are important for controlling ozone depletion in the stratosphere, and winds and small scale mixing. Extensive detail about the atmospheric modeling, including a full discussion of all included reactions and processes, as well as discussion of modeling uncertainties, can be found in Thomas et al. (2005) and Ejzak et al. (2007).

Melott and Thomas (2009) reviewed a wide range of events and concluded that supernovae, short duration hard spectrum gamma-ray bursts (SHGRBs), and long duration soft spectrum gamma-ray bursts (LSGRBs) (respectively) are the most likely astrophysical events to pose serious threats to life on Earth over a few 100 million year time scales. Piran and Jimenez (2014) arrive at similar conclusions for LSGRBs.

Ejzak et al. (2007) showed that atmospheric changes following astrophysical radiation events depend on total energy and spectral hardness, not duration of the event. All such radiation events result in ionization of the atmosphere, primarily in the stratosphere and above, which leads to production of odd-nitrogen compounds (usually denoted $NO_y$) and subsequent depletion of stratospheric $O_3$. For this study, the modeled change in $O_3$ is the parameter of interest; therefore, the exact astrophysical event chosen is not important. Our results may be applied to any event with similar atmospheric consequences (determined solely by event hardness and energy fluence), including SNe, GRBs and extreme SPEs.

We have chosen to focus our analysis here on the results of one particular case: a GRB of "long" duration (10 seconds) and "soft" spectrum (Band photon spectrum with $E_0 = 187.5$ keV) with fluence 100 kJ m$^{-2}$, occurring over Earth's South Pole, in late June (roughly the Southern Hemisphere winter solstice). This choice was motivated by several factors. First, Melott and Thomas (2009) have argued that a South-Polar burst fits well with what is currently known about the late Ordovician mass-extinction. In brief, modeled biological damage as a function of latitude for that case matches well with measured extinction rate, peaking around 35° South latitude. Second, we wished to examine an extreme (but still realistic) event, with severe stratospheric $O_3$ depletion, leading to higher production of low altitude $O_3$; a polar burst isolates the ozone-depleting compounds to that hemisphere, which tends to result in greater localized $O_3$ depletion/production. The maximum localized decrease in total vertical column density of $O_3$ for this case is about 70%, and the maximum globally averaged decrease in total vertical column density is about 32% (Thomas et al., 2005), roughly the level of depletion assumed to be mass-extinction level in previous work (e.g. Melott and Thomas, 2011).

Thomas et al. (2005) modeled cases of a GRB occurring over five different latitudes and at four different times during the year (the equinoxes and solstices); differences in intensity and time/location distribution of $O_3$ depletion are discussed extensively there. For the present work we examined ground-level $O_3$ for all of these cases. While increases relative to the "background" of an unperturbed run vary in amount and time/location distribution between cases, we find only small variation in maximum absolute values of ground-level $O_3$, which is the relevant measure for biological impact considered in this work. The South-Polar, June event shows the largest absolute values, which again motivates our focus on that case.

We searched the literature for data on how organisms are affected by $O_3$ and what levels lead to significant damage. A significant body of work exists on the effects on humans and crop plants. As noted above, the late Ordovician extinction has been identified as a good candidate for being connected to a GRB. There was very little terrestrial life at that time and the extinction is recorded primarily in marine organisms. Therefore, we also examine the impact of increased $O_3$ on marine organisms. While astrophysical ionizing radiation events are only likely to affect Earth on geologic timescales, we use the existing literature as a guide to what levels of $O_3$ are likely to be harmful to organisms. Though there was little land life at the late Ordovician, our results may be applied to a range of ionizing radiation events, not just GRBs, and so impacts on terrestrial plants and animals is relevant for studies of events such as supernovae, which are likely to have occurred at other times during the Phanerozoic, on roughly 300-500 million year time scales (Melott and Thomas, 2011).

3. Results

Here we present results of our analysis of previously completed atmospheric modeling following a GRB, as described above. Figure 1 shows the percent difference in $O_3$ concentration between the GRB model run and a control run (no GRB input) in the lowest altitude bin of the atmospheric chemistry model, which extends from the ground to approximately 2 km in altitude. We show the full latitude range over a duration of 5 years after the event, which occurred on June 21 in the model run (indicated by time 0 here). Note that since the event we consider was modeled to occur over the South Pole, the atmospheric effects are essentially limited to the Southern hemisphere. This is due to the fact that radiation from the GRB produces $NO_y$ in the middle atmosphere, where transport is primarily pole-ward. Chemistry changes, therefore, are most extreme around the South Pole in the case considered here, and we restrict our analysis to the Southern Hemisphere from this point on.

Note in Figure 1 that there is an annual cycle of increase and decease in ground-level $O_3$ in the model, most noticeable at Southern latitudes greater than about 60°, following the seasonal cycle of increasing and decreasing sunlight. Change in ground-level $O_3$ in our modeling is controlled primarily by penetration of Solar UV to much lower altitudes than normal due to extreme depletion of $O_3$ in the stratosphere and higher. Figure 2 shows percent difference in $O_3$ concentration at 65° South latitude, as a function of altitude (from the ground to about 50 km) and time. Note the extreme depletion in the middle and upper stratosphere immediately following the event. As ozone-depleting $NO_y$ compounds are transported downward, especially during Polar winter, $O_3$ concentrations begin to return to normal in the upper stratosphere but remain depleted in the lower stratosphere. (Note that in these results Southern hemisphere winter solstice occurs at months 0, 12, etc.) When depletion of $O_3$ at middle and high altitudes is greatest, production of $O_3$ in the lower atmosphere is greatest. The seasonal cycle in $O_3$ change at altitudes below about 10 km results from a similar cycle at higher altitudes, combined with increased descent of $NO_y$ compounds during the winter months, when downward transport within the South Polar vortex region is strongest (see Figure 3).

The seasonal depletion in $O_3$ (compared to the control run) in the lowest altitude bin of the model at the most extreme Southern latitudes is caused by several factors. First, stratospheric ozone depletion is strongest during South Polar winter, but the sun angle is low (or sunlight is absent entirely) so that low altitude production of $O_3$ is minimal. Second, in the control run (normal atmospheric conditions), some $O_3$ from the stratosphere is transported downward to lower altitudes in the South Polar vortex; however, in the GRB run depletion of stratospheric $O_3$ means less $O_3$ arrives at lower altitudes by this transport. (Full analysis of these dynamics is outside the scope of this work, since our purpose here is to evaluate the absolute values of $O_3$ at the surface in the GRB case.) Finally, $NO_y$ compounds are transported downward within the South Polar vortex during this time, thereby causing some depletion of $O_3$ at lower altitudes. This effect explains the increasing intensity of low altitude $O_3$ depletion seen in Figure 1. Overall, then, several factors interact to lead to the reductions in $O_3$ compared to the control run in the lowest altitude bin of the model, at the highest Southern latitudes, during South Polar winter.

In Figure 4 we show the $O_3$ concentration in parts per billion (ppb) at the lowest altitude bin of the atmospheric model in the GRB case. This plot shows only the latitude region where $O_3$ in the lowest altitude bin in the model is increased following the GRB (see Figure 1), since we wish to only examine the $O_3$ levels at the ground as increased by the GRB. We find a maximum of about 10 ppb $O_3$ in the lowest 2 km in the model results for this region, varying seasonally, as noted above.

We now consider what level of $O_3$ concentration may be considered hazardous to organisms on Earth's surface. While astrophysical ionizing radiation events are only likely to affect Earth on geologic timescales, we may use the existing literature as a guide to what levels of $O_3$ are likely to be harmful to organisms. An important factor to note is that exposure time is important to consider, as well as absolute values of $O_3$ concentration, since longer term exposure leads to more biological damage. In our modeling, ground-level $O_3$ is enhanced seasonally for about 6 months at a time, over the first three years after the event (see Figure 1).

According to the United States Environmental Protection Agency's Health Risk and Exposure Assessment (United States Environmental Protection Agency, 2014) there are adverse health risks to young healthy adults with exposure at 72 ppb of ozone for 6.6 hours. The EPA has recently proposed to set the National Ambient Air Quality Standards for ground-level ozone to a value in the range 65 – 70 ppb, with suggestions that the value should be set at 60 ppb. Jerrett et al. (2009) report a long term study (1977-2000) of ozone exposure and mortality in humans. They find an enhanced risk of death due to respiratory causes in areas with $O_3$ concentration above about 30 ppb, with an increase in risk of death of about 2.9% for every 10 ppb increase in ozone.

Long term (months to years) exposure to ground-level ozone damages vegetation, with substantial reduction in crop yields and crop quality (Ghude et al., 2014). As a strong oxidant, ozone (or secondary products resulting from oxidation by ozone such as reactive oxygen species) causes several types of symptoms including chlorosis (discoloration due to reduced chlorophyll production) and necrosis (cell death).  The severity of the injury is dependent on several factors including duration and concentration of ozone exposure, weather conditions and plant genetics (Bell and Treshow, 2002).

Studies of ozone's influence on crop yields differ in their results. Studies of soybean yield at the University of Maryland found a 10 percent loss of soybean crop due to current levels of ozone in that state, which are commonly 40-80 ppb during the growing season, with particular episodes much higher. The same study showed that ozone exposure causes the loss of 6-8 percent of winter wheat and 5 percent of the corn crop yields to Maryland farmers (Maryland Department of the Environment, 2013).

Several studies have quantified yield losses on a global or regional scale (Heagle, 1989; Wang and Mauzerall, 2004; Holland et al., 2006; Aunan et al., 2000).  A threshold value of 40 ppb has been used in impact assessment research in Europe (Mills et al., 2007), while a higher value of 60 ppb has been used in the USA. There is evidence that ozone may affect vegetation at concentrations well below 40 ppb and a lower threshold of 30 ppb would be more appropriate.

Given the literature discussed above, we may consider a value of 30 ppb as our threshold for harmful effects on both vegetation and animals, using humans as a model.  As noted above (see Figure 4), the maximum $O_3$ concentration in regions affected by the GRB in our modeling is about 10 ppb.  Therefore, the amount of ground-level ozone associated with our modeled GRB is too small to be likely to have a major effect on vegetation or (using humans as a model) animals.

The biological impact of $O_3$ dissolved in sea water is also relevant, since over geologic time life has existed longer and with more diversity in Earth's oceans, and during the Ordovician period life was almost wholly restricted to the oceans. Ozone is highly soluble in water; about ten times more soluble than oxygen, and has been used to destroy microorganisms in water since the early 20$^{th}$ century (Sonntag and Gunten, 2012). Reduction of active cells to $10^{-4}$ of the original number can be achieved by one minute exposure at concentrations of $1.2 \times 10^{-3}$ mg L$^{-1}$ for the bacterium *E. coli*, and 11.8 mg L$^{-1}$ for the protozoa *C. parvum* (Sonntag and Gunten, 2012). According to Asbury and Coler (1980), a variety of fish species larvae are sensitive to residual $O_3$ concentrations of about $5.0 \times 10^{-2}$ mg L$^{-1}$. The 10 ppb maximum $O_3$ concentration in air discussed above corresponds to about $10^{-5}$ mg L$^{-1}$. Even if all of this $O_3$ became dissolved in the sea water the concentration would be 2 orders of magnitude smaller than that needed to adversely affect *E. coli* in one minute exposures, and 3 orders of magnitude smaller than that tolerable by fish larvae indefinitely (residual concentration). In our scenario exposures much longer than one minute would occur. Therefore, we may expect some impact on the most sensitive organisms (ie. those similar to *E. coli*), but little if any impact on most aquatic species.

4. Discussion and Conclusions

There is little doubt that astrophysical ionizing radiation events have the potential to cause significant impact to Earth's biosphere, for a range of energy fluence values. The main impact is increased Solar UV at the surface following stratospheric ozone depletion. Previously neglected is the fact that ozone is actually increased in the lower atmosphere for several years following such events. In this work we considered the changes in ground-level ozone for the case of a GRB using previously completed atmospheric chemistry modeling. We found that while there is an increase in ozone in the lowest 2 km altitude bin of the model, the modeled concentration is too low to significantly impact organisms on the ground or in the oceans. Given that the GRB case considered here represents one of the most energetic astrophysical ionizing radiation events likely to affect life on Earth over geologic time periods, we may safely rule out ground-level ozone increases as an additional biological hazard for ionizing radiation events. This conclusion applies to other events of similar or lesser fluence and/or softer spectra (Ejzak et al. 2007; Melott and Thomas, 2011), including SNe and extreme SPEs, on any planet with an atmosphere similar to that of modern Earth.

Acknowledgements

The authors thank the two anonymous reviewers for helpful comments that significantly improved the manuscript. This work has been supported by the National Aeronautics and Space Administration under grant No. NNX14AK22G, through the Astrobiology: Exobiology and Evolutionary Biology Program. Computational time for this work was provided by the High Performance Computing Environment (HiPACE) at Washburn University; thanks to Steve Black for assistance with computing resources.

Author Disclosure Statement: No competing financial interests exist.

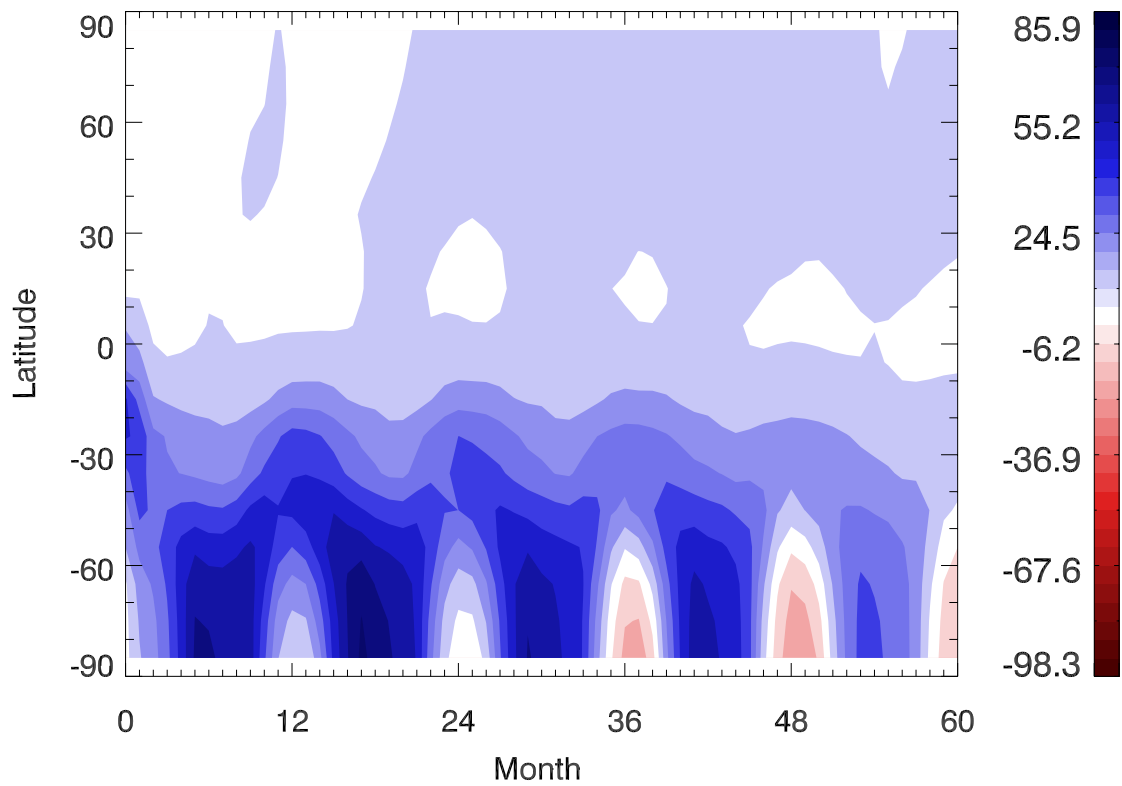

Figure 1

Percent difference (between a run with GRB ionization input and a control run) in $O_3$ number density at the lowest altitude bin of the atmospheric chemistry model.

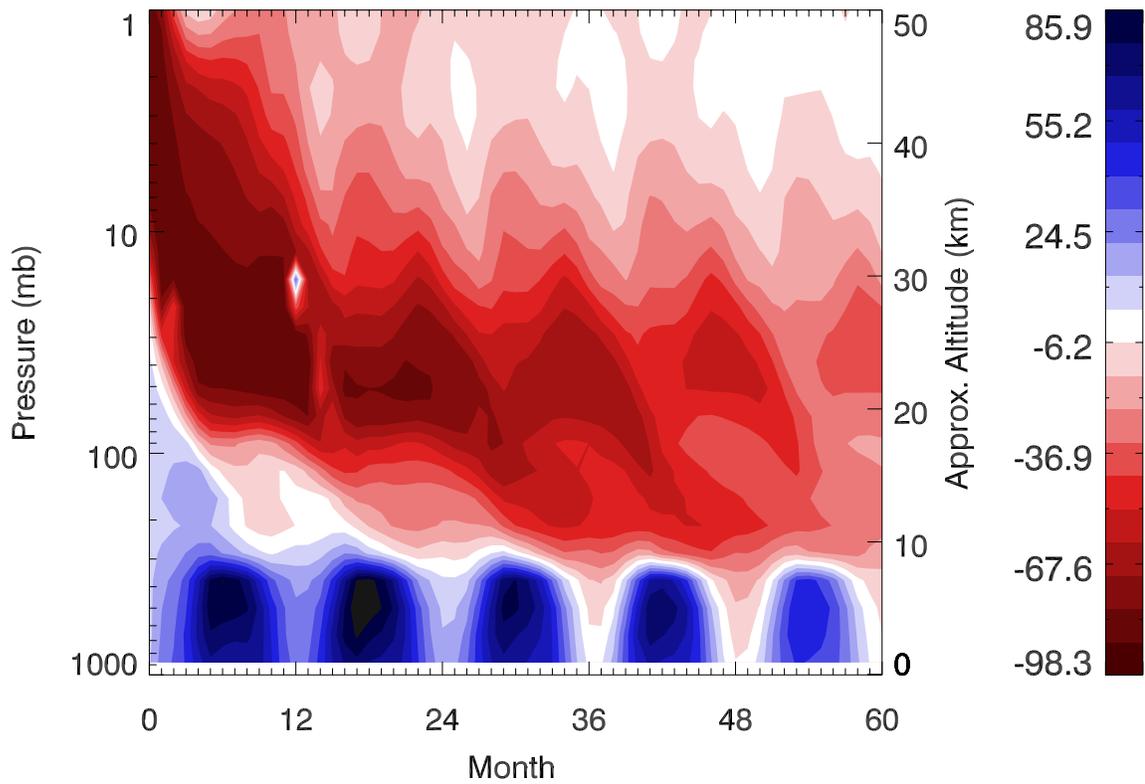

Figure 2

Percent difference in $O_3$ number density at 65° South latitude as a function of pressure (approximate altitude), from the ground to the top of the stratosphere, and time.

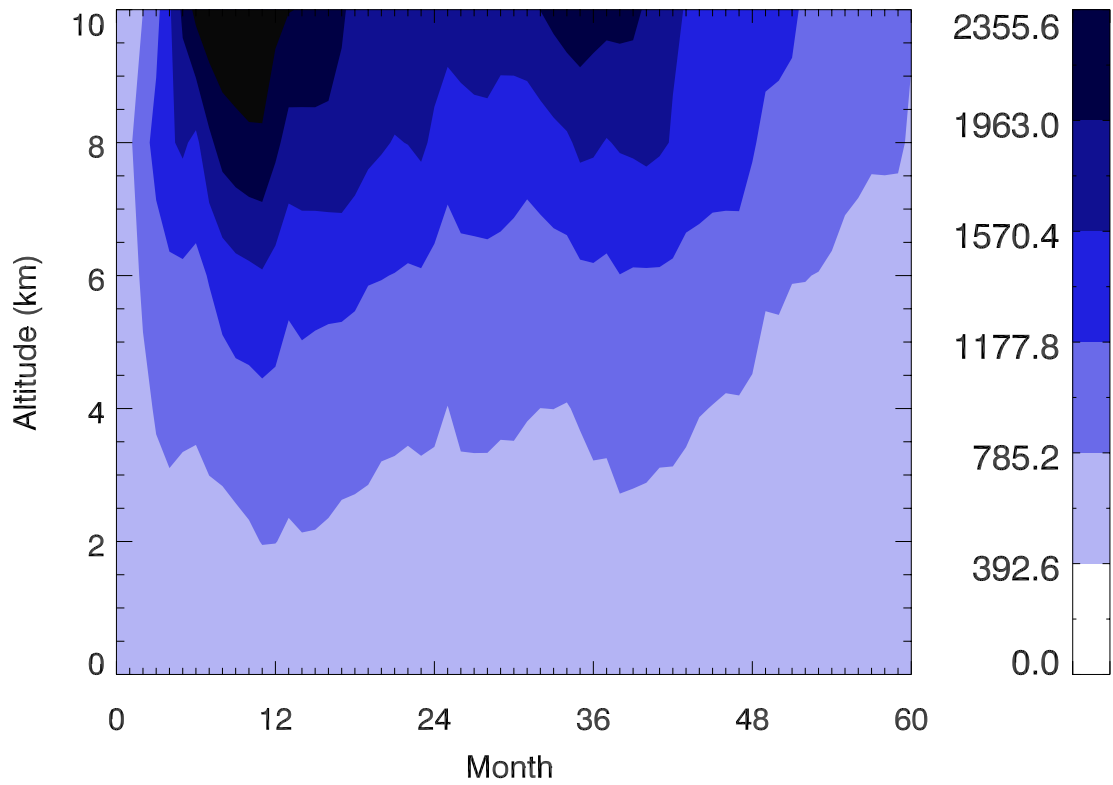

Figure 3

Percent difference in $NO_y$ number density at 65° South latitude as a function of altitude, from the ground to approximately the top of the troposphere, and time.

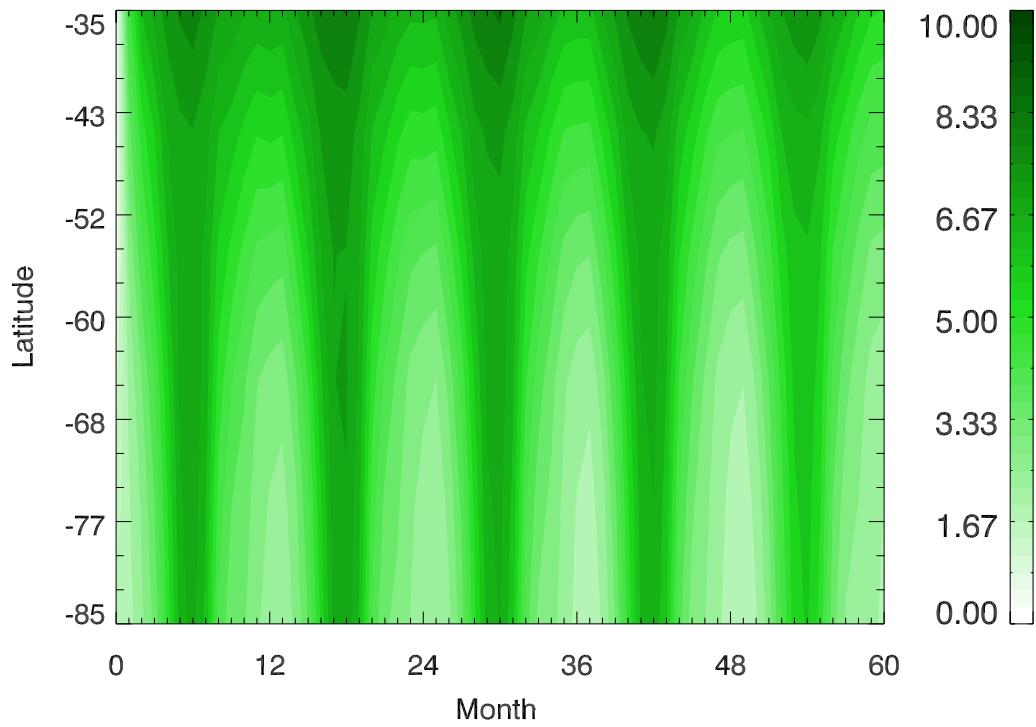

Figure 4

Concentration of $O_3$ (in ppb) in the lowest altitude bin of the atmospheric chemistry model. The latitude range is chosen to coincide with the area most strongly affected by the modeled event (see Figure 1).